# Multi-layer Optimized Coordination of Smart Building Resources in Active Power Distribution Systems

Mohammadali Rostami, *Member, IEEE,* Saeed Lotfifard, *Senior Member, IEEE,* Mladen Kezunovic, *Life Fellow, IEEE*

***Abstract*—This paper proposes a multi-actor coordination platform for the optimal utilization of smart buildings resources, including roof top PV generation and battery energy storage system (BESS), in active power distribution systems. The proposed multi-actor coordination includes the Smart Building Coordinator (SBC), Micro-Grid Coordinator (MGC) and Distribution System Coordinator (DSC). The coordinators operate independently and only exchange limited information with each other to reach an optimal solution. In the proposed platform, a hierarchical optimization problem is solved to optimally determine the operating point of all distribution system resources. The proposed platform fully preserves the confidentiality of the behind the meter (BTM) data of the buildings since no information about the status of the PV system, BESS, and load of the building is shared with the owner of the power system. The proposed platform has a flexible and scalable architecture where the computational task of coordinating microgrids and smart buildings with distribution grid is performed locally at the MGC and SBC layers, respectively. Numerical simulations show the efficacy of the proposed platform in coordinating the BTM resources with the rest of the distribution system.**

## NOMENCLATURE

| | |
|---|---|
| $C_{DG}$ | Generation cost of DG units |
| $E$ | State of the charge of BESS |
| $E_0$ | Initial state of the charge of BESS |
| $f^*$ | Value of the objective function in ATC algorithm |
| $i, j$ | Subscripts denoting bus numbers |
| $k$ | Iteration counter in ATC |
| $min$, $max$ | Superscripts denoting maximum and minimum of variables |
| $N_{dist}$ | Total number of buses in distribution network |
| $N_{MG}$ | Total number of buses in MG |
| $o$ | Vector product operator in ATC |
| $P_B$, $Q_B$ | Exchanged active and reactive power between SB and grid |
| $P_d$, $Q_d$ | Active and reactive power demand at a node |
| $P_{DG}$, $Q_{DG}$ | Active and reactive power generation of DG units |
| $P_{ESS}$ | Injected/absorbed active power by BESS |
| $P_g$, $Q_g$ | Active and reactive power generation at a node |
| $P_{Grid}$ | Exchanged active power between distribution system and upstream network |
| $P_{ij}$,$Q_{ij}$ | Active and reactive power flow between buses $i$ and $j$ |
| $P_{MG}$, $Q_{MG}$ | Active and reactive power exchange between MG and distribution system |
| $P_{PV}$ | Generated active power by PV panel |
| $P_{PV}^{forecast}$ | Predicted solar active power |
| $P_L$ | Total load demand of SB |
| $r$ | Vector of response variables in ATC |
| $S$ | Subsystems in ATC |
| $S_{B,dist}$ | Set of all buses with SB in distribution network |
| $S_{B,MG}$ | Set of all buses with SB in MG |
| $S_{DG,dist}$ | Set of all buses with DG units in distribution network |
| $S_{DG,MG}$ | Set of all buses with DG units in MG |
| $S_{ij}^{max}$ | Maximum apparent power flow between bus $i$ and $j$ |
| $S_{MG}$ | Set of boundary buses between MG and distribution system |
| $S_{line,dist}$ | Set of all feeders in distribution network |
| $S_{line,MG}$ | Set of all feeders in MG |
| $t$ | Vector of target variables in ATC |
| $V$ | Voltage magnitude of buses |
| $w_1,\ldots,w_9$ | Weighting factors used to normalize objective functions |
| $Y_{ij}\angle\theta_{ij}$ | Admittance between buses i and j |
| $\alpha$, $\beta$ | vector of coefficients for linear and quadratic parts of the penalty function in ATC |
| $\delta$ | Voltage angle of buses |
| $\varepsilon_1$, $\varepsilon_2$ | Convergence tolerance in ATC |
| $\eta$ | Charging/discharging efficiency of the BESS |
| $\pi$ | augmented Lagrangian penalty function in ATC |
| $\rho$ | Electricity price |
| $\sigma$ | Convergence acceleration multiplier in ATC |
| $\lambda$ | Penalizing coefficient for exchanged power between MG and distribution network |

M. Rostami and S. Lotfifard are with the School of Electrical Engineering and Computer Science, Washington State University, Pullman, WA, 99164, USA (e-mail: mohammadali.rostami@wsu.edu, s.lotfifard@wsu.edu).

M. Kezunovic is with the Department of Electrical and Computer Engineering, Texas A&M University, College Station, TX 77840 USA (e-mail: kezunov@ece.tamu.edu)

## I. INTRODUCTION

EFFICIENT operation of smart buildings connected to power distribution systems can lead to improved energy efficiency as buildings make up a large segment of total energy consumption [1]. Rooftop solar system and battery energy storage system (BESS) are installed at the commercial and residential buildings enabling bidirectional power exchange between buildings and distribution systems. This paradigm shift in distribution system interfacing leads to new operational and control challenges that necessitate a coordinated operation of smart building (SBs) resources. Different methods and control strategies have been proposed for energy management of buildings [2], [3]. In [4] a resource allocation problem is solved for PV and storage units in buildings. An architecture is proposed for integrating buildings with rooftop PV in smart grids in [5]. In [6] a demand side management strategy that considers the presence of smart buildings in the grid is proposed. An energy management strategy for thermal and air conditioning resources of SBs is proposed in [7]. An energy management system for residential buildings based on fuzzy logic is designed in [8]. An optimal control approach for the air conditioning systems of the building is designed in [9]. A smart energy management scheme to reduce the overall energy consumption of the buildings is developed in [10]. In [11] a building energy management system (BEMS) that coordinates the operation of the building resources is proposed. These models. The interactive load management of buildings in smart grids is discussed in [12]. A cooperative power control model among a network of residential buildings that are connected to the main grid is developed in [13]. The operation of integrated residential loads in smart grid environment is optimized in [14]. In [15] a building-to-grid (B2G) energy management strategy that models the residential buildings as active loads in the network is proposed. The integration and coordination of the renewable energy resources in the buildings is studied in [16], [17].

In summary, the previous research can be categorized as follows: a) methods that only consider the unidirectional flow of power from the grid to the buildings and manage the SB energy without considering the impacts of power system such as [9]-[11]; b) methods that consider the bidirectional power exchange between SBs and the grid where the distribution network model is simplified or not considered at all [12]- [17]; and c) methods for B2G (building to grid) energy management, which consider the distribution network model [18] but the sustainability of the buildings is not addressed, and passive distribution network model is used while the presence of DGs and MGs are not considered.

In this paper, our contribution is the formulation of the optimal utilization of SB resources in coordination with the active distribution systems as a tri-layered optimization framework. We proposed an architecture platform shown in Fig.1. It demonstrates that the proposed platform fully preserves the confidentiality of the BTM data of SBs because in the proposed platform SBs do not need to share any information about the status of the BTM resources with the rest of the power system. The only shared information between the SBs and the grid is the net exchanged power. The advantage of this feature is that it provides more incentives for building resource owners to participate in energy transactions with active distribution systems. Also, the proposed framework enhances the scalability of the energy management systems of the active distribution system as it enables the building operators to perform the required computations locally with minimum information exchange with the distribution system operator.

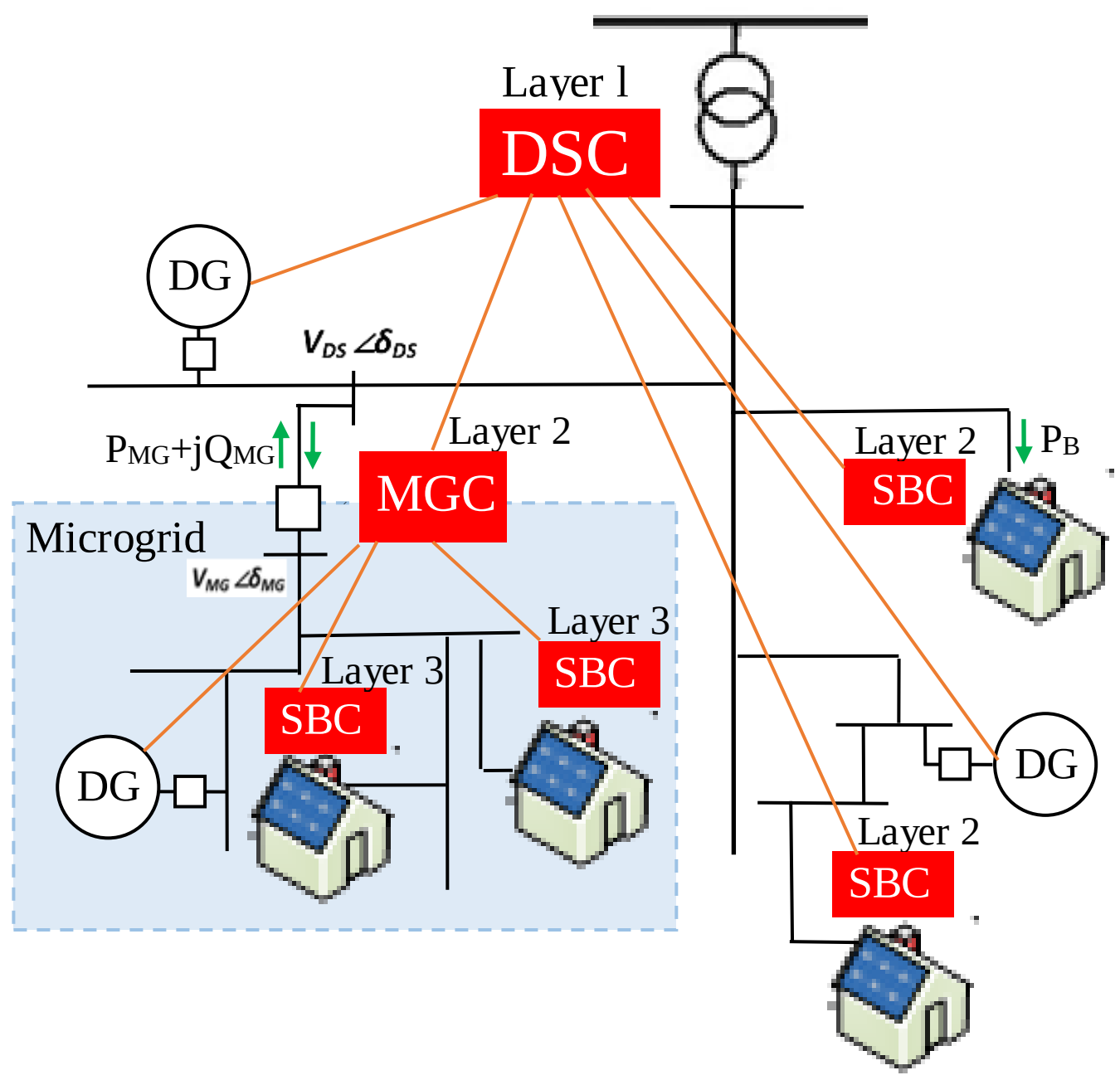

Fig. 1. Schematic of the proposed multi-actor coordination platform comprising Smart Building Coordinator (SBC), Micro-Grid Coordinator (MGC) and Distribution System Coordinator (DSC)

The proposed platform has a flexible architecture in which MG and SB coordinators can operate in grid-connected or islanded modes.

The rest of this paper is organized as follows: In section II the hierarchical architecture for managing smart buildings in distribution system is described. The mathematical formulation of the multi-actor coordinated optimization model and the solution procedure is presented in Section III. Numerical simulations and results discussions are provided in Section IV, and the paper is concluded in Section V.

## II. Multi-Actor Architecture for Managing Smart Buildings Interfaced to Active Distribution Systems

Fig.1 shows the proposed multi-actor architecture for operating active power distribution systems. The DSC operator exchanges limited information with operators of MGC and SBCs that are directly connected to the grid (i.e. SBs that are not part of any MGs). SBs are directly connected to either the distribution system network or to MGs which is able to operate in both grid-connected and islanded modes [19]-[21]. MGs can be utility-owned or privately-owned subsystems. Each MGC exchanges information with the DSC at the higher layer and various SBCs that are located within the MG at the lower layer. The SBCs located within MSGs constitute the third layer information exchange. SBC operators decide whether to store the generated power by the roof top PV panels in BESS or trade power with the grid to meet a set of predefined requirements. The information exchange among the operators can be performed via variety of communication systems [22]-[24].

## III. Coordinated Optimization Formulation

In this section the optimization problem at DSC, MGC, and SBC is formulated and the procedure to coordinate solution of the multi-actor optimization problem is presented. The objective is to balance the power flows in power distribution systems and is solved for the next hour. The transition of MGs to islanded mode is also anticipated in the proposed formulation in both DSC optimization problem and MGC optimization problem. This transition becomes smoother if the exchanged power between MG and distribution grid is reduced. This can be considered by adding a term to the objective function of DSC and MGC optimization problems in order to reduce the exchanged power between distribution system and MG.

### A. Optimization Formulation for Distribution System

Distribution system exchanges power with the upstream transmission network, MGs and SBs that are directly connected to the distribution network (i.e. SBs that are not part of any MGs). The DSC minimizes the operating costs and power exchange with MGs while meeting the demand and operational requirements of the network.

The operation optimization at DSC is formulated as follows:

$$\min. \quad \omega_1 \rho P_{Grid} + \omega_2 \sum_{i \in S_{DG,dist}} C_{DG,i} P_{DG,i} + \omega_3 \sum_{i \in S_{B,dist}} \rho P_{B,i} + \omega_4 \sum_{i \in S_{MG}} \lambda ({P_{MG,i}}^2 + {Q_{MG,i}}^2) \quad (1)$$

Subject to

$$P_{inj,i} = \sum_{j=1}^{N_{dist}} V_i V_j Y_{ij} \cos(\delta_i - \delta_j - \theta_{ij}) \quad i = 1 : N_{dist} \quad (2)$$

$$Q_{inj,i} = \sum_{j=1}^{N_{dist}} V_i V_j Y_{ij} \sin(\delta_i - \delta_j - \theta_{ij}) \quad i = 1 : N_{dist} \quad (3)$$

$$P_{inj,i} = P_{g,i} - P_{d,i} \quad i \notin S_{B,dist} \quad (4)$$

$$Q_{inj,i} = Q_{g,i} - Q_{d,i} \quad i \notin S_{B,dist} \quad (5)$$

$$P_{inj,i} = -P_{B,i} \quad i \in S_{B,dist} \quad (6)$$

$$Q_{inj,i} = -Q_{B,i} \quad i \in S_{B,dist} \quad (7)$$

$$V_i^{\min} \le V_i \le V_i^{\max} \quad i = 1 : N_{dist} \quad (8)$$

$$P_{ij}^2 + Q_{ij}^2 \le (S_{ij}^{\max})^2 \quad ij \in S_{line,dist} \quad (9)$$

$$P_{DG,i}^{\min} \le P_{DG,i} \le P_{DG,i}^{\max} \quad i \in S_{DG,dist} \quad (10)$$

$$Q_{DG,i}^{\min} \le Q_{DG,i} \le Q_{DG,i}^{\max} \quad i \in S_{DG,dist} \quad (11)$$

$$P_{B,i}^{\min} \le P_{B,i} \le P_{B,i}^{\max} \quad i \in S_{B,dist} \quad (12)$$

in which, $P_{Grid}$ is the exchanged power between distribution system and upstream transmission network. $C_{DG}$ is the generation cost of DG units and $\rho$ is the electricity price. $\lambda$ is the coefficient factor for the exchanged power between MGs and distribution network. $P_g$ and $Q_g$ are the generated active and reactive power at network buses respectively. $P_{DG}$ and $Q_{DG}$ are the generated active and reactive power by DG units respectively. $P_d$ and $Q_d$ are the total active and reactive power demand at system buses respectively. $P_{ij}$ and $Q_{ij}$ denote the active and reactive line flows between buses $i$ and $j$ respectively. $P_B$ represents the provided power from the grid to the SB which is limited by $P_B^{min}$ sand $P_B^{max}$. Note that negative value of $P_B$ means the SB injects power into the grid. $Q_B$ is the reactive power exchange of SB with the grid and it is determined based on SB power factor. In this paper a unity power factor is assumed for the SB, i.e. $Q_B$=0. $P_{MG}$ and $Q_{MG}$ are the exchanged active and reactive power between distribution system and MG respectively. $V_i \angle \delta_i$ is the bus voltage at $i$th bus and $Y_{ij} \angle \theta_{ij}$ is the admittance between buses $i$ and $j$. The minimum and maximum limits of bus voltage magnitudes are denoted by $V^{max}$ and $V^{min}$ respectively. The limits on active power generation of DGs are denoted by $P_{DG}^{min}$ and $P_{DG}^{max}$. Similarly, the limits on reactive power generation of DG units are given by $Q_{DG}^{min}$ and $Q_{DG}^{max}$. The maximum apparent flow on the lines is denoted by $S^{max}$. The coefficients $w_1$, $w_2$, $w_3$, and $w_4$ are the weighting factors to normalize the objective function. The weight factors are constant values which can be calculated as the initial value of each term [25]. Finally, the set of SBs, DG units and MGs, in the distribution network are denoted by $S_{B,dist}$, $S_{DG,dist}$, and $S_{MG}$ respectively.

DSC optimization is a nonlinear problem because the power flow constraints are nonlinear. The main decision variables are $P_{Grid}$, $P_{DG}$, $P_B$, $P_{MG}$, $Q_{MG}$, $V$, $\delta$, $P_{ij}$ ,and $Q_{ij}$. The objective function (1) consists of four terms. The first term is the cost of buying power from upstream transmission network, the second term is the cost of power from DGs, and the third term is the cost of power from SB resources. The role of this term is to maximize the profit of the SB resource owners to incentivize further deployment of SBs. The last term aims at minimizing the amount of exchanged power with MGs. A quadratic term is used since the power exchange between distribution system and MGs is bidirectional. As explained previously, minimizing the power exchange between distribution system and MGs facilitates a smoother transition of MGs to islanded mode in the case of emergency situations. There is no need to consider this term during normal operation of network which is simply achieved by setting $w_4$=0. Constraints (2)-(7) represent the power flow equations. The bus voltages and line flows are limited by constraints (8)-(9) respectively. The active and reactive power generation of DGs are restricted by constraints (10)-(11) respectively. The exchanged power between the SBs and distribution grid is limited by constraint (12).

### B. Optimization Formulation for MGs

MGCs interact with the DSC and SBCs that are located within the MGs. Each MGC solves the following optimization problem:

$$\min. \quad \omega_5 \sum_{i \in S_{DG,MG}} C_{DG,i} P_{DG,i} + \omega_6 \sum_{i \in S_{B,MG}} \rho P_{B,i} + \omega_7 \lambda ({P_{MG}}^2 + {Q_{MG}}^2) \quad (13)$$

Subject to

$$P_{inj,i} = \sum_{j=1}^{N} V_i V_j Y_{ij} \cos(\delta_i - \delta_j - \theta_{ij}) \quad i = 1 : N_{MG} \quad (14)$$

$$Q_{inj,i} = \sum_{j=1}^{N} V_i V_j Y_{ij} \sin(\delta_i - \delta_j - \theta_{ij}) \quad i = 1 : N_{MG} \quad (15)$$

$$P_{inj,i} = P_{g,i} - P_{d,i} \quad i \notin S_{B,MG} \quad (16)$$

$$Q_{inj,i} = Q_{g,i} - Q_{d,i} \quad i \notin S_{B,MG} \quad (17)$$

$$P_{inj,i} = -P_{B,i} \quad i \in S_{B,MG} \quad (18)$$

$$Q_{inj,i} = -Q_{B,i} \quad i \in S_{B,MG} \tag{19}$$

$$V_i^{\min} \le V_i \le V_i^{\max} \quad i = 1: N_{MG} \tag{20}$$

$$P_{ij}^{\,2} + Q_{ij}^{\,2} \le (S_{ij}^{\max})^2 \quad ij \in S_{line,MG} \tag{21}$$

$$P_{DG,i}^{\min} \le P_{DG,i} \le P_{DG,i}^{\max} \quad i \in S_{DG,MG} \tag{22}$$

$$Q_{DG,i}^{\min} \le Q_{DG,i} \le Q_{DG,i}^{\max} \quad i \in S_{DG,MG} \tag{23}$$

$$P_{B,i}^{\min} \le P_{B,i} \le P_{B,i}^{\max} \quad i \in S_{B,MG} \tag{24}$$

MGC optimization is also a nonlinear problem because of power flow constraints. In the above formulation, the set of SBs and DG units in MG are denoted by $S_{B,MG}$ and $S_{DG,MG}$ respectively. The main decision variables in this optimization problem are $P_{DG}$, $Q_{DG}$, $P_B$, $P_{MG}$, $Q_{MG}$, $V$, $\delta$, $P_{ij}$ ,and $Q_{ij}$. The objective function (13) consists of three terms which are weighted by normalizing coefficients $w_5$, $w_6$, and $w_7$. The weighting factors are constant values which are calculated as the initial value of each term [25]. The first term denotes the cost of power generation by DGs and the second term is the power cost for buildings in the MG. The last term is considered to reduce the amount of exchanged power between the $i$th MG and distribution system. As explained previously, this term facilitates smoother transition of an MG to islanded mode in case of emergency and there is no need to consider this term under normal conditions. This term can be ignored by simply setting $w_7$=0. Constraints (14)-(19) are the nodal power flow equations at each bus, and constraint (20) is the limits on bus voltage magnitudes. Constraints (21) limits the power flow on the network feeders. Constraints (22)-(23) are active and reactive generation of DG units respectively. Finally, the exchanged power between the SBs and MG is limited according to (24).

### C. *Optimization Formulation for SB Resources*

The buildings are equipped with roof top PV systems and BESS that supply power to their loads as depicted in Fig.2. The overall demand of the building includes both controllable and non-controllable loads. SBC aims at minimizing the operational cost by coordinating the operation of roof top PV systems, BESS, and power exchange with the power grid. The power from the grid could be used to supply the load or stored in the storage device to be used at the later time. The total operating cost of SB includes the cost of power exchange between SB and the grid to which the building is connected to. The building is either connected to the distribution system or MG. Mathematically, the following optimization problem is solved by the $i$th SB:

$$\min. \quad \rho P_{B,i} \tag{25}$$

Subject to

$$P_{B,i} + P_{PV,i} + P_{ESS,i} = P_{L,i} \tag{26}$$

$$E_i = E_{i,0} + \eta_i P_{ESS,i} \tag{27}$$

$$E_i^{\min} \le E_i \le E_i^{\max} \tag{28}$$

$$P_{ESS,i}^{\min} \le P_{ESS,i} \le P_{ESS,i}^{\max} \tag{29}$$

$$P_{PV,i} \le P_{PV,i}^{forecast} \tag{30}$$

$$P_{L,i}^{\min} \le P_{L,i} \le P_{L,i}^{\max} \tag{31}$$

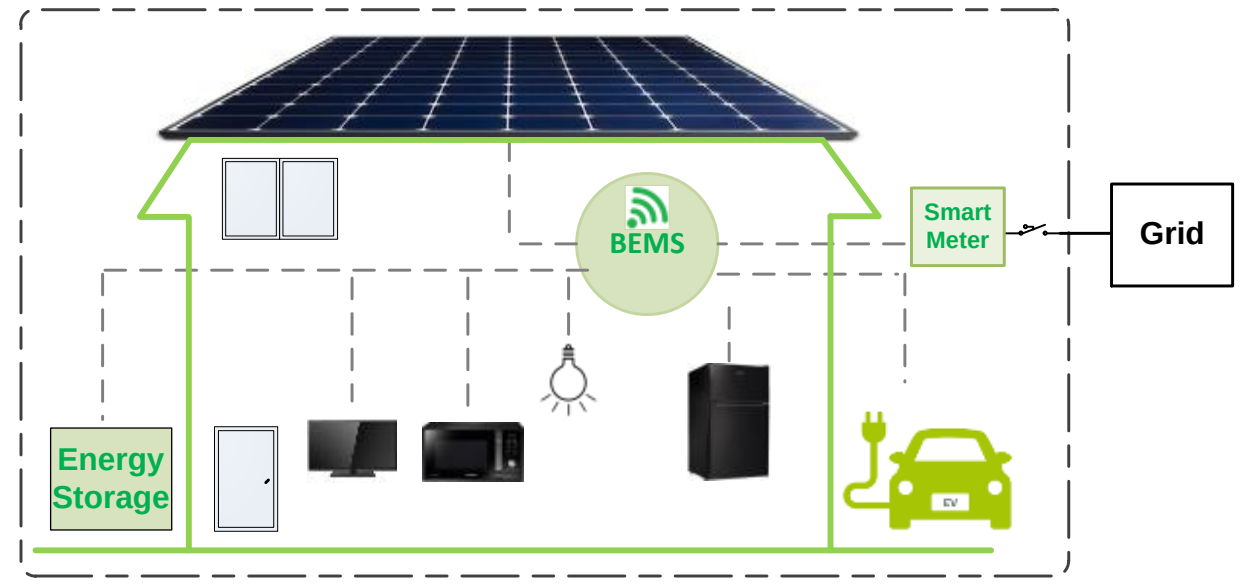

Fig. 2. Conceptual smart building structure

$$P_{B,i}^{\min} \le P_{B,i} \le P_{B,i}^{\max} \tag{32}$$

where $P_{PV}$ is the generated power by solar PV panel, $P_{ESS}$ is the power provided by the storage unit, and $P_L$ is the total load demand of the building. $E$ is the state of charge of the BESS and $\eta$ is the charging/discharging efficiency of the BESS [4]. The current state of the charge of the BESS is noted by $E_0$. The minimum and maximum limits on state of charge of the ESS are denoted by $E^{min}$ and $E^{max}$ respectively. $P_{PV}{}^{forecast}$ is the predicted solar generation. $P_L{}^{min}$ and $P_L{}^{max}$ denote the limits on load demand of the building.

SBC optimization problem is a linear optimization problem in which the objective function and all the constraints are linear. In the SB operation optimization, the main decision variables are $P_B$, $P_{ESS}$, $E_i$, $P_{PV}$, and $P_L$. The objective function reflects the cost or revenue from exchanging power between SB and upstream grid. The power exchange between SB and the network is bidirectional since the SB is equipped with storage devices, roof top PV generation and controllable load. At convenient time, it might be more beneficial for the building to turn off its controllable load and sell the excess power to the grid. Constraint (26) assures the power balance between different sources and total load demand in the building. Constraint (27) is the energy balance in the storage device. The energy capacity of the storage device is limited by constraint (28). The continuity of the power supply to SBs in the case of outages in distribution systems is assured by increasing the limit on the minimum stored energy. The storage device is permitted to exchange power with the grid such that its remaining energy would be sufficient for the SB critical demand for a specified time plus some extra amount due to minimum allowable SOC. The charging and discharging rates of the storage device are restricted by constraints (29). The solar PV generation is limited by the forecasted value according to constraint (30). Finally, the total load and the amount of exchanged power between the grid and SB is limited by constraints (31)-(32) respectively. As discussed earlier, the total load of the building ($P_L$) consists of uncontrollable loads and controllable loads which may be curtailed to minimize the total operating cost of the building.

### D. *Coordinated optimization procedure*

The optimization problems described for each layer in previous subsections are coordinated using the analytical target cascading (ATC) method [26]-[27].

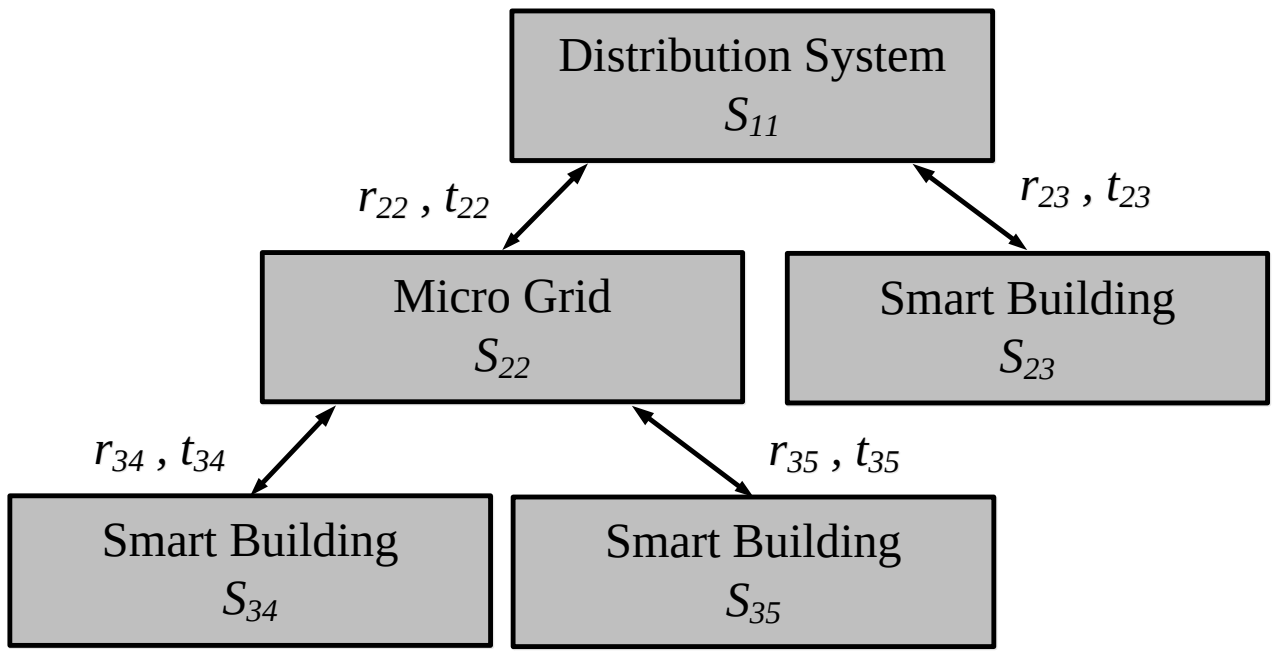

Fig. 3. Coordinated distribution system interaction

The multi-actor architecture of a generic distribution system with MGs and SBs is illustrated in Fig. 3.

In the example structure shown in Fig. 3, one SB and one MG are directly connected to the distribution system.

Two SBs are also directly connected to the MG. The distribution system ($S_{11}$) is the first layer, MG ($S_{22}$) and SB ($S_{23}$) are in the second layer, and SBs ($S_{34}$ and $S_{35}$) are in the third layer. For each of these subsystems the optimization problem is formulated as explained in previous sections. The optimization problem for the jth subsystem in the ith layer can be written in the following general form:

$$\min . f_{ij}(X_{ij}) \tag{33}$$

Subject to

$$g_{ij}(X_{ij}) \le 0 \tag{34}$$

In which $f$ is the objective function, $g$ represents the constraints, and $X$ is the vector of decision variables. The objective function, constraints and decision variables are described for each optimization problem in the previous section. For each subsystem the set of response variables and target variables should also be determined. The target and response variables are the shared variables between two connected subsystems. According to Fig. 1, the shared variables between the distribution system and MG are the voltage of boundary buses, i.e. $\{V_{DS}, V_{MG}, \delta_{DS}, \delta_{MG}\}$, and the shared variable between SBs and the distribution grid is the exchanged power, i.e. $P_B$. Therefore, in DSC optimization problem, the vector of target variables are $t=\{V_{DS}, V_{MG}, \delta_{DS}, \delta_{MG}, P_B\}$ which are the decision variables determined by DSC optimization problem. Also, the vector of response variables are $r=\{V_{DS}^*, V_{MG}^*, \delta_{DS}^*, \delta_{MG}^*, P_B^*\}$. The sign * denotes the values pertaining to the solution of the optimization problem given as constant values by the lower layers. The response variables are determined by MGC optimization problem and SBC optimization problem in the lower level, and are added to the DSC optimization problem as fixed values using the following penalty function [28]:

$$\pi_{ij} = \alpha_{ij}^T \,|\, t_{ij} - r_{ij} \,| + \| \beta_{ij} \circ (t_{ij} - r_{ij}) \|_2^2 \tag{35}$$

in which, $t_{ij}$ and $r_{ij}$ denote the vector of target and response variables of the subsystem $j$ in layer $i$ respectively where $\alpha_{ij}$ and $\beta_{ij}$ are the vector of coefficients of the penalty function and operator $o$ represents the vector product where elements with identical elements in vector are multiplied. Therefore, DSC solves the following optimization problem:

**Algorithm 1.** ATC algorithm for multi-actor SB management

1: **Initialize** all target and response variables by performing the power flow analysis in the base case system, iteration counter ($k$=1), and algorithm parameters including coefficients of penalty functions.

2: Solve **SB optimization problem** for all SBs with their respective penalty function.

3: Solve **MG optimization problem** for all MGs with calculated $P_B$ from previous step as input.

4: Solve **distribution system optimization** problem with updated boundary voltages from step 3 and updated $P_B$ values from step 2.

5: Check **convergence criteria**. If they are not satisfied proceed to next step, otherwise, stop the algorithm and report optimal solutions. Both criteria should be satisfied for convergence:

$$\left| \frac{f_{ij}^{*(k)} - f_{ij}^{*(k-1)}}{f_{ij}^{*(k)}} \right| \le \varepsilon_1 \tag{38}$$

$$t_{ij}^* - r_{ij}^* \le \varepsilon_2 \tag{39}$$

6: set $k$=$k$+1, and **update coefficients** of penalty functions according to following equations:

$$\alpha_{ij}^{(k+1)} = \alpha_{ij}^{(k)} + 2(\beta_{ij}^{(k)})^2 (t_{ij}^{*(k)} - r_{ij}^{*(k)}) \tag{40}$$

$$\beta_{ij}^{(k+1)} = \sigma \beta_{ij}^{(k)} \tag{41}$$

Updating coefficients increases the weight of the penalty function and forces the target and response variables to become closer.

$$\min . f_{ij}(X_{ij}) + \pi_{ij} \tag{36}$$

Subject to

$$g_{ij}(X_{ij}) \le 0 \tag{37}$$

in which $f_{ij}$ is the objective function in (1), $g_{ij}$ are constraints (2)-(12), and $\pi_{ij}$ is (35). For DSC of generic system in Fig. 3, $t_{ij}=\{t_{22}, t_{23}\}$ because it is connected to one MG and one SB. $t_{22}=\{V_{DS}, V_{MG}, \delta_{DS}, \delta_{MG}\}$ and $t_{23}=\{P_B\}$ are decision variables and $r_{22}=\{V_{DS}^*, V_{MG}^*, \delta_{DS}^*, \delta_{MG}^*\}$, $t_{23}=\{P_B^*\}$ are calculated values by MG and SB respectively. The subscripts $DS$ and $MG$ denote the boundary buses between distribution grid and MG as depicted in Fig. 1. Once $t_{22}$, $t_{23}$ are calculated by DSC, they are passed as constant values to MGC and SBC respectively.

Similarly, MG solve (36)-(37) in the second layer where $f_{ij}$ is the objective function in (13), $g_{ij}$ are constraints in (14)-(24), and $\pi_{ij}$ is (35). For MGC $t_{ij}=\{t_{34}, t_{35}\}$ because it is connected to two SBs in third layer. $t_{34}$ and $t_{35}$ are the $P_B$ of respective SBs and they are both decision variables for MGC. $r_{ij}=\{r_{22}\}=\{V_{DS}^*, V_{MG}^*, \delta_{DS}^*, \delta_{MG}^*\}$ are the solutions calculated by DSC and passed to MGC as constant values. Once $t_{34}$ and $t_{35}$ are calculated by MGC they are passed to SBs as constant values.

For the SB in the second layer ($S_{23}$ in Fig. 3), SBC solves (36)-(37) where $f_{ij}$ is objective function in (25), $g_{ij}$ are constraints in (25)-(32), and $\pi_{ij}$ is (35). $t_{ij}=\{t_{23}\}=\{P_B\}$ is the decision variable for SBC and $r_{ij}=r_{23}=\{P_B^*\}$ is constant value calculated by DSC. Once $t_{23}$ is calculated by SBC, it is passed to DSC as constant values.

Finally, for SBs in the third layer ($S_{34}$ and $S_{35}$ in Fig. 3), SBC solves (36)-(37) where $f_{ij}$ is objective function in (25), $g_{ij}$ are constraints in (25)-(32), and $\pi_{ij}$ is (35). $t_{ij}$={$t_{23}$}={$P_B$} is the decision variable for SBCs and $r_{ij}$=$r_{23}$={$P_B^*$} is the constant value calculated by DSC. Once $r_{23}$ is calculated by SBC, it is passed to DSC as constant values.

The above approach is repeated iteratively to achieve convergence as described in Algorithm 1. Algorithm 1 outlines the ATC approach for the proposed multi-actor optimization problem. In Algorithm 1, $f_{ij}^*$ is the current (iteration $k$) value of objective function for subsystem $j$ in layer $i$. $t_{ij}^*$ and $r_{ij}^*$ are the current target and response solutions respectively. $\varepsilon_1$ and $\varepsilon_2$ are desired convergence tolerances. $\sigma\geq1$ is a multiplier to speed up the convergence.

In summary, in ATC the subsystem optimization problems are solved at the upper layer to determine the target variables which are then propagated to lower layer subsystems. The optimization of lower layer subsystem operation uses these target variables as inputs to determine the response variable which are propagated back to the higher-layer subsystems. The propagation of target and response variables between the upper layer and lower layer subsystems are repeated iteratively until the desired coordination among subsystems is achieved. The consistency penalty function assure that target and response variables are in coordination between upper later and lower layer subsystems. More explanations and mathematical derivations on convergence properties of ATC method can be found in [29]-[30].

## IV. Simulation Results

The numerical simulations are performed on the 33-bus distribution test system. The network data can be found in [31]. The distribution network includes two MGs as depicted in Fig. 4. The distribution system interacts with the upstream transmission network, two MGs, two DG units, and two SBs. All the optimization problems are solved using the solvers provided by *opti* [32].

Each MG exchanges power with the distribution system as well as the DG unit and SB within the MG. The total active and reactive load of the network is 3.40 MW and 2.11 MVAr respectively. Four 0.8 MW DG units are installed in the network. MG1 is connected to the distribution system at bus two and MG2 is connected to the distribution network at bus six. Therefore, the shared variables between MG1 and the distribution system are {$V_2$, $V_{19}$, $\delta_2$, $\delta_{19}$}and the shared variables between MG2 and distribution system are {$V_6$, $V_{26}$, $\delta_6$, $\delta_{26}$}. The shared variables between the SBs and the network are the exchanged power between the SBs and the network at their connecting bus. The specifics of the buildings are given in Table I. The controllable loads constitute 30 percent of the total load of SBs and the predicted solar generation for each SB is 3.2 kW. The convergence tolerances for the ATC algorithm is set to 0.001 for both $\varepsilon_1$ and $\varepsilon_2$. The minimum and maximum limits of bus voltages is set to 0.95 p.u. and 1.05 p.u. respectively.

Two different cases are simulated in order to study the performance of the proposed tri-layer coordination platform.

*Case 1*: Network operation under normal situation

*Case 2*: Network operation anticipating islanding of MG

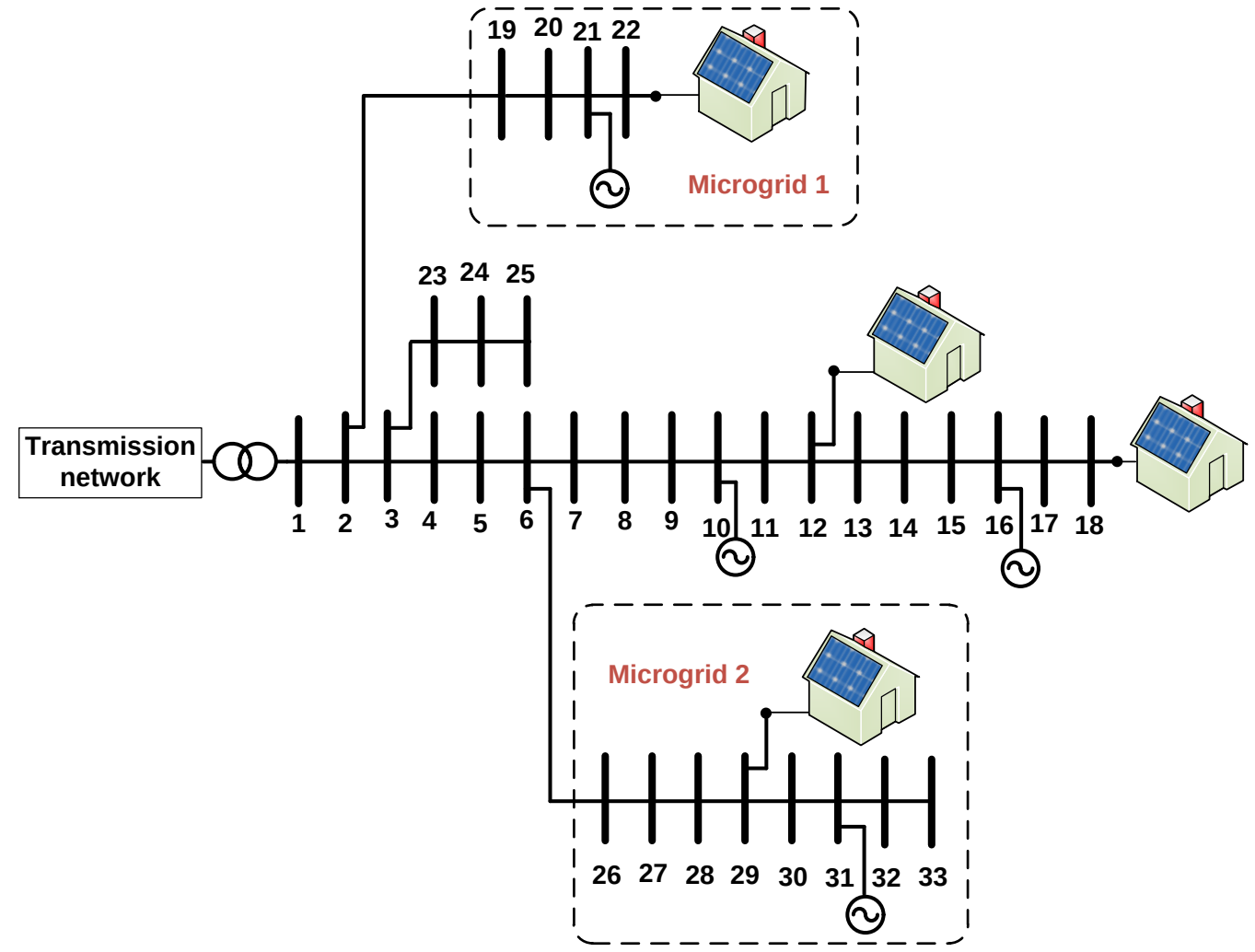

Fig. 4. Schematics of the simulated test system

Table I. Specifics of SBs in the network

| Bus Number | 12 | 18 | 22 | 29 |
|---|---|---|---|---|
| Total Load (kW) | 10 | 9 | 8 | 12 |
| ESS Capacity (kWh) | 3 | 5 | 5.5 | 8 |
| Charge/Discharge Rate (kW/h) | 1.5 | 1.8 | 1.7 | 1.6 |

Table II. SBs dispatches

| | Case 1 | | | | Case 2 | | | |
|---|---|---|---|---|---|---|---|---|
| Bus Number | 12 | 18 | 22 | 29 | 12 | 18 | 22 | 29 |
| $P_L$ (kW) | 7 | 6.3 | 5.6 | 8.4 | 7 | 6.3 | 5.6 | 8.4 |
| $P_{ESS}$ (kW) | 0.79 | 1.32 | 1.45 | 1.6 | 0.79 | 1.32 | 1.45 | 1.6 |

Table III. Active and reactive generation of DGs

| | Case 1 | | | | Case 2 | | | |
|---|---|---|---|---|---|---|---|---|
| Bus Number | 10 | 16 | 21 | 31 | 10 | 16 | 21 | 31 |
| $P_{DG}$ (MW) | 0.77 | 0.08 | 0.8 | 0.19 | 0.50 | 0.42 | 0.53 | 0.47 |
| $Q_{DG}$ (MVAr) | 0.39 | -0.18 | -0.34 | 0.4 | 0.32 | 0.05 | -0.21 | 0.27 |

Table IV. Exchanged power with transmission network

| | Case 1 | Case 2 |
|---|---|---|
| $P$(MW) | -0.39 | 3.40 |
| $Q$ (MVAr) | 0.78 | 1.43 |

In case 1, limitations on the power exchange between the MGs and the distribution system are ignored. Therefore, $\omega_4$ and $\omega_7$ are set to zero in DSC and MGC optimization problems. In case 2, islanded operation for the MGs is expected mainly due to anticipated adverse weather condition and the limiting term on power exchange between distribution system and MG is considered in both objective functions of DSC and MGC optimization problems.

The dispatch of SBs and generation of DG units is compared in Table II – III for both cases. The amount of active and reactive power exchanged with the transmission network is given in Table IV for both cases.

The bus voltages and angles are given in Table V. The results show that the exchanged power between the distribution network and MG1 is reduced from 0.79 MVA to 0.41 MVA in the second case. The exchanged power between the distribution network and MG2 is reduced from 0.80 MVA to 0.15 MVA in the second case. According to the results, the proposed framework has been successful in finding the operating point of the power system elements in both cases. As expected, the amount of exchanged power between MG and distribution grid is reduced in case 2 where the islanded operation of MG is anticipated. Moreover, according to Table V, the bus voltages in both cases are within the provided secure limits (0.95 p.u $\leq V \leq$ 1.05 p.u).

Table V. bus voltages and angles of the network

| | *V* (pu) | | *δ* (rad) | |
|---|---|---|---|---|
| Bus No. | Case1 | Case 2 | Case 1 | Case 2 |
| 1 | 0.9982 | 1.0057 | 0 | 0 |
| 2 | 0.9978 | 1.0034 | -0.0001 | -0.0002 |
| 3 | 0.9926 | 0.9970 | -0.0008 | -0.0014 |
| 4 | 0.9918 | 0.9951 | -0.0013 | -0.0022 |
| 5 | 0.9912 | 0.9936 | -0.0019 | -0.0031 |
| 6 | 0.9908 | 0.9909 | -0.0036 | -0.0062 |
| 7 | 0.9906 | 0.9902 | -0.0023 | -0.0082 |
| 8 | 0.9926 | 0.9889 | -0.0014 | -0.0090 |
| 9 | 0.9973 | 0.9890 | 0.0013 | -0.0107 |
| 10 | 1.0025 | 0.9896 | 0.0042 | -0.0122 |
| 11 | 1.0026 | 0.9895 | 0.0045 | -0.0120 |
| 12 | 1.0027 | 0.9895 | 0.0050 | -0.0116 |
| 13 | 1.0026 | 0.9889 | 0.0078 | -0.0097 |
| 14 | 1.0026 | 0.9888 | 0.0092 | -0.0088 |
| 15 | 1.0035 | 0.9896 | 0.0105 | -0.0078 |
| 16 | 1.0028 | 0.9889 | 0.0103 | -0.0081 |
| 17 | 1.0021 | 0.9881 | 0.0098 | -0.0086 |
| 18 | 1.0021 | 0.9881 | 0.0098 | -0.0086 |
| 19 | 0.9989 | 1.0019 | 0.0005 | -0.0002 |
| 20 | 1.0048 | 0.9980 | 0.0056 | 0.0044 |
| 21 | 1.0067 | 0.9969 | 0.0075 | 0.0059 |
| 22 | 1.0067 | 0.9969 | 0.0075 | 0.0059 |
| 23 | 0.9825 | 0.9868 | -0.0014 | -0.0019 |
| 24 | 0.9792 | 0.9836 | -0.0029 | -0.0034 |
| 25 | 0.9904 | 0.9909 | -0.0036 | -0.0041 |
| 26 | 0.9903 | 0.9907 | -0.0037 | -0.0060 |
| 27 | 0.9894 | 0.9902 | -0.0030 | -0.0053 |
| 28 | 0.9855 | 0.9875 | -0.0008 | -0.0024 |
| 29 | 0.9830 | 0.9860 | 0.0010 | -0.0001 |
| 30 | 0.9819 | 0.9855 | 0.0023 | 0.0013 |
| 31 | 0.9835 | 0.9882 | 0.0020 | 0.0017 |
| 32 | 0.9826 | 0.9873 | 0.0016 | 0.0014 |
| 33 | 0.9824 | 0.9871 | 0.0015 | 0.0013 |

## V. Conclusion

A tri-layer optimization coordination platform was proposed for managing active distribution grids. It offers several advantages:

- Offers optimal management of smart building resources considering the interactions between active distribution systems and smart buildings equipped with BESSs and roof top PV generation.
- Calculates optimized operation cost of distribution systems and SBs while enhancing the continuity of electricity supply to SBs.
- Each subsystem (i.e. SBs and MGs) solves its own optimization problem locally with limited information exchanges according to ATC.
- Enables SBs to preserve confidentiality of the BTM resources as they do not need to share the status of the BTM data such as PV panel output power, energy storage state of charge, and the consumer electricity use with the rest of the power system operators.
- Does not require the privately owned microgrid owners to share information of the corresponding subsystems with the distribution systems operators.
- Demonstrates the effectiveness of the proposed multi-actor platform for coordinating the operation of SBs with the rest of active distribution system.
- It can be used as the foundation for solving more complex problems such as studying the communication between different layers during dynamic system behavior and variable electricity price at different elements of the system.

## VI. BIOGRAPHIES

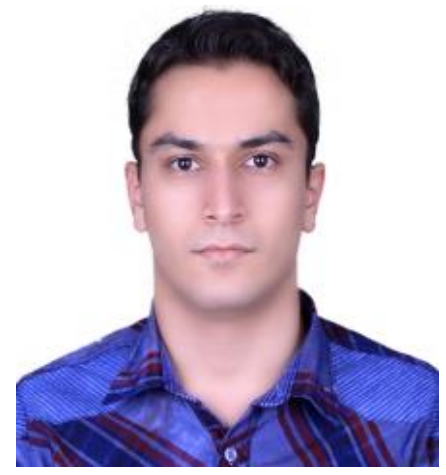

**Mohammadali Rostami** (S'15-M'20) received his PhD degree in electrical engineering from Washington State University, Pullman, in 2019. His research interest includes wide area monitoring, protection and control of power systems.

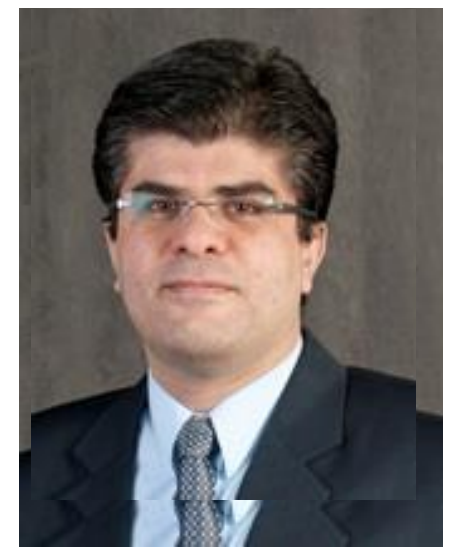

**Saeed Lotfifard** (S'08–M'11-SM'17) received his Ph.D. degree in electrical engineering from Texas A&M University, College Station, TX, in 2011. Currently, he is an associate professor at Washington State University, Pullman. His research interests include stability, protection and control of inverter-based power systems. Dr. Lotfifard is an associate editor for the IEEE Transactions on Power Delivery.

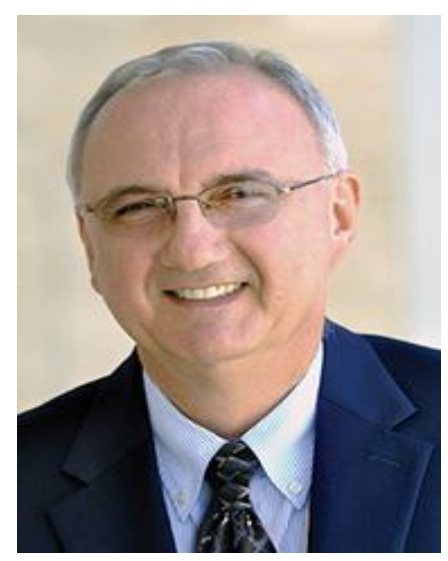

Mladen Kezunovic (S'77–M'80–SM'85–F'99–LF'17) received the Dipl. Ing., M.S., and Ph.D. degrees in electrical engineering in 1974, 1977, and1980, respectively. He has been with Texas A&M University since 1986 where he is currently Distinguished Professor, Regents Professor, Wofford Cain Chair IV, and the Site Director of Power Engineering Research Center. His expertise is in protective relaying, automated power system disturbance analysis, computational intelligence, data analytics, and smart grids. He has published over 650 papers, given over 200 seminars, invited lectures and short courses, and consulted for over 80 companies worldwide. He is the Principal of XpertPower Associates, a consulting firm specializing in power systems data analytics. Dr. Kezunovic is a CIGRE Fellow, Honorary Member and Distinguished Member, Registered Professional Engineer in TX, USA and Member of the National Academy of Engineering.